\documentclass[11pt]{article}
\usepackage{amssymb,amsmath,mathrsfs,enumerate}
\usepackage{graphicx,rotate,multicol,wrapfig}
\usepackage[margin=10pt,labelfont=bf]{caption}
\usepackage{cite}
\usepackage[utf8]{inputenc}
\usepackage[colorlinks=true,
            linkcolor=red,
            urlcolor=blue,
        citecolor=blue]{hyperref}

\usepackage{color}
\usepackage{lineno}
\long\def\rpl#1!!!#2!!!{\color[rgb]{.7,0,0}{#1} \color{blue}{#2} \color{black}}

 \usepackage[normalem]{ulem}

 \definecolor{darkgreen}{cmyk}{1,0,1,0.4}

\let\bar=\overline

\def \order(#1){{\mathcal O} \left(#1 \right)}

\evensidemargin=\oddsidemargin
\def\Eqn#1{Eq.\ (\ref{#1})}
\def\Eqs#1#2{Eqs.\ (\ref{#1}) and (\ref{#2})}
\allowdisplaybreaks

\title	{\Large\bf 
Higgs data does not rule out a sequential fourth generation  \\
with an extended scalar sector
}

\author {\sf Dipankar Das,$^{a,}$\footnote{ddphy@caluniv.ac.in} \quad  
Anirban Kundu,$^{a,}$\footnote{anirban.kundu.cu@gmail.com} \quad 
Ipsita Saha$^{b,}$\footnote{ipsita.saha@roma1.infn.it} \\[10pt]
\small\em $^a$Department of Physics, University of Calcutta, 92
Acharya Prafulla Chandra Road, Kolkata 700009, India\\ 
\small\em $^b$Instituto Nazionale di Fisica Nucleare, Sezione di Roma,
Piazzale Aldo Moro 2, I-00185 Roma, Italy \\ 
 }

\date{}

\begin{document}


\maketitle	

\begin{abstract}
Contrary to the common perception, we show that the current Higgs data does not eliminate the possibility
of a sequential fourth generation that get their masses through the same Higgs mechanism as the 
first three generations. The inability to fix the sign of the bottom-quark Yukawa coupling from the
available data plays a crucial role in accommodating a chiral fourth generation which is consistent
with the bounds on the Higgs signal strengths.
 We show that effects of such a fourth generation can remain 
completely hidden not only in the production of the Higgs boson through gluon fusion but also to its
subsequent decay to $\gamma\gamma$ and $ Z\gamma$. 
This, however, is feasible only if the scalar sector of the Standard Model is extended.
We also provide a practical example illustrating how our general prescription can be embedded in a realistic model.
\end{abstract}

\bigskip


\paragraph{Introduction:}
Since the first observation of a new resonance at the Large Hadron Collider~(LHC)
\cite{Aad:2012tfa,Chatrchyan:2012xdj} in 2012, the particle physics community
converged on the view that this  is a 
Standard Model~(SM)-like Higgs boson if not the SM Higgs boson itself. 
The fact that the LHC Higgs data is gradually drifting towards the SM 
expectations\cite{Khachatryan:2016vau}
has pushed many scenarios that go beyond the SM~(BSM) to some contrived corner of the
parameter space. For some BSM, the situation is even worse as they have nowhere
to hide, because quantum effects coming from some of the heavy degrees of freedom
do not decouple and hence leave observable imprints
in the Higgs signal strengths. The SM extended by a chiral fourth generation~(SM4)\cite{Kribs:2007nz} constitutes
such an example.
The fourth generation quark masses, so heavy as to avoid the direct detection bound, 
are proportional to the corresponding Yukawa couplings and thus, their contributions
to the $gg\to h$ production amplitude saturate to a constant value just as in the case
of the top quark loop. Consequently, the gluon-gluon fusion~(ggF) amplitude for the Higgs
production increase roughly by a factor of 3 compared to the SM, enhancing the cross-section
by a factor of 9. This should have been reflected as a huge enhancement in the Higgs
signal strengths, nothing like which has been observed, leading us to believe that the possibility
of a sequential fourth generation is strongly disfavored from the existing data \cite{Eberhardt:2012gv,
Djouadi:2012ae,Vysotsky:2013gfa,Kuflik:2012ai,Lenz:2013iha}.

Efforts have been made to counter the enhanced production in the ggF channel by reducing the
branching ratios~(BRs) of the Higgs boson into different visible channels. This can be achieved, {\em e.g.},
by adjusting the mass of the fourth generation neutrino~($m_{\nu'}$) so that the Higgs boson
mainly decays invisibly into a pair of fourth generation neutrinos thereby increasing the total
decay width of the Higgs boson\cite{Khoze:2001ug,Belotsky:2002ym,Cetin:2011fp,Englert:2011us,CuhadarDonszelmann:2008jp,Carpenter:2010dt,Keung:2011zc,Aparici:2012vx,Geller:2012tg}. However this possibility also fell out of favor
after the arrival of the Higgs data\cite{Kuflik:2012ai}.
Another way to avoid the enhancement in the ggF production channel
 is to assume that the heavy chiral fourth generation
receives its masses from a different scalar other than the SM Higgs doublet\footnote{
Although, at the initial stage of the Higgs discovery when the data was not so precise,
 the possibility of the fourth generation coupling to the SM Higgs doublet could have been
entertained in different BSM scenarios 
 \cite{Bellantoni:2012ag,Banerjee:2013hxa}, these options became increasingly disfavored
 by the LHC Higgs data constantly evolving towards the SM expectations.}. 
 This option has been considered in the framework of a two Higgs doublet model~(2HDM)
 \cite{BarShalom:2011zj,BarShalom:2012ms,He:2011ti,Chamorro-Solano:2017toq}. 
 
In this article, we ask a more ambitious question: Can we still accommodate an extra
generation of fermions coupling to the SM Higgs {\em in an identical way} as the first three generations,
without altering appreciably the Higgs signal strengths from their corresponding SM expectations?
We will answer in the affirmative and provide a practical example of such a scenario.
As we will explain, this is feasible because the Yukawa couplings of up- and down-type quarks can be 
so arranged as to make the extra contributions cancel out. Such a cancellation,
quite remarkably, is also effective
for loop-induced Higgs decays. However, a realization of this is not possible with only one scalar 
doublet,\footnote{
One can make all the Yukawa couplings positive by suitable chiral rotations for a single
scalar doublet coupling to all the fermions.}
so one needs to extend the scalar sector of the SM.

\paragraph{Making the fourth generation survive:}
The way to make a fourth generation survive is to note the subtle fact that the sign of the bottom quark 
Yukawa coupling is not determined from the present LHC data. 
The sign of the top quark Yukawa coupling has to be positive with respect to the $WWh$ coupling (where $h$ is
the SM, or SM-like, Higgs boson with a mass of 125 GeV), which can be inferred 
from the interference in
 the $h\to \gamma\gamma$ amplitude. The mass being small, effect of the bottom quark loop is 
negligible as far as current precision is concerned.

Let us now define the Higgs coupling modification factors relative to the SM as follows:
\begin{eqnarray}
\kappa_x =\frac{g_{xxh}}{\left(g_{xxh}\right)_{\rm SM}} \,,
\end{eqnarray}
where, $x=W,Z$ and other massive fermions, and $g_{xxh}$ is the generic coupling. 
For the SM, all $\kappa_x=1$, which is of course completely consistent with the LHC data. However, as 
the tree-level two-body decays of $h$ cannot shed any light on the phase of the coupling, one may also 
entertain an alternative scenario, namely,
\begin{subequations}
\label{e:ws}
\begin{eqnarray}
\kappa_V = 1 && (V=W,Z) \\
\kappa_u = 1 && ({\rm for~ up~ type~ quarks}) \\
\kappa_d = -1 && ({\rm for~ down~ type~ quarks~ and ~charged ~leptons})\,.
\end{eqnarray}
\end{subequations}
In what follows, we will refer to this possibility as the {\em wrong sign limit}.

The modification factor for the $gg\to h$ production cross section in
the presence of fourth generation quarks is given by
\begin{eqnarray}
R_{gg} &=& \frac{\left|\kappa_tF_{1/2}(\tau_t)+\sum\limits_{f=t',b'} \kappa_fF_{1/2}(\tau_f) \right|^2}{\left|F_{1/2}(\tau_t)\right|^2}
\end{eqnarray}
where, using $\tau_x \equiv (2m_x/m_h)^2$, the expression for $F_{1/2}$
is given by \cite{Gunion:1989we}
\begin{eqnarray}
F_{1/2}(\tau_x) = -2\tau_x \big[1+(1-\tau_x)f(\tau_x)\big] \,.
\end{eqnarray}
Since we are concerned with heavy fermions, we can take
$\tau_{x} > 1$ for $x = (t,t',b')$, and then
\begin{equation}
f(\tau) =\left[\sin^{-1}\left(\sqrt{1/\tau}\right)\right]^2\,.
\end{equation}
 For chiral fermions much heavier than $m_h=125$~GeV, the loop function $F_{1/2}$ saturates to a constant
value \cite{Gunion:1989we} and the new physics~(NP) contribution  becomes proportional
to $(\kappa_{t'}+\kappa_{b'})$. Clearly, in the SM-like limit ($\kappa_{t'}=\kappa_{b'}=1$),
$R_{gg}=9$, which dealt a killer blow to the sequential fourth generation within the SM. 
But, more importantly, in the wrong sign limit, the NP contributions 
coming from the $t'$ and $b'$ loops are of opposite sign and hence there exists the possibility that 
they may cancel each other. Strictly speaking, the cancellation is perfect in the limit $m_{t'}=m_{b'}$, but even if there is a 
mass splitting, it is still exact for all intents and purposes as long as $t'$ and $b'$ are much heavier than $h$. 
Thus, there is no enhancement in Higgs production through gluon-gluon fusion.

At this point, one might suspect that such a perfect cancellation might not occur for
the $h\to\gamma\gamma$ decay. But we should remember that, for anomaly
cancellation \cite{Adler:1969gk,Bell:1969ts}, one needs to include an extra chiral generation of leptons too.
The fourth generation charged lepton, $\tau'$, also contributes to the diphoton
decay. Consequently, the NP contribution to the $h\to\gamma\gamma$ amplitude,
in the heavy mass limit, is proportional to
\begin{eqnarray}
\label{e:gaga}
\kappa_{\gamma\gamma} = \sum\limits_{f=t',b',\tau'}
Q_f^2 \, N_c^f\, \kappa_f \,,
\end{eqnarray}
where, $Q_f$ is the electric charge of the fermion $f$, 
and $N_c=3$ for quarks and $1$ for leptons. One can
easily check that $\kappa_{\gamma\gamma} =0$ in the wrong sign limit. Thus a heavy chiral
extra generation can remain perfectly hidden from the LHC Higgs data in this  limit. 
In passing, we note that, the quantity
\begin{eqnarray}
\label{e:kZga}
\kappa_{Z\gamma} = \sum\limits_{f=t',b',\tau'}
Q_f T_3^f N_c^f\kappa_f \,,
\end{eqnarray}
where $T_3^f$ denotes the isospin projection of $f_L$, also vanishes in the wrong sign limit
leaving no trace of extra generations in the $h\to Z\gamma$ decay as well. The mechanism, obviously, works even for two
or more such heavy generations as well.

\paragraph{A practical example:}
The reader, at this point, might wonder whether the conspiracy of couplings given in \Eqn{e:ws} can be
realized in a gauge theoretic model. The
 wrong sign limit is not achievable in the
 SM itself with only one scalar doublet,
 because of the uniform proportionality between the fermion masses and their Yukawa 
couplings. Also, one may note that the cancellation of the bad high-energy behaviour for
 the $f\bar{f}\to W_LW_L$
scattering amplitude is sensitive to the relative sign between the $WWh$ and $f \bar{f} h$
 couplings \cite{Bhattacharyya:2012tj}. Consequently, in the wrong sign limit, the amplitude at high energies for
$d\bar{d}\to W_LW_L$ (where $d$ represents a generic down-type quark or a charged lepton)
will grow and eventually violate the partial-wave unitarity,
 if only one Higgs doublet is present in the theory.
Hence, we must extend the scalar sector.

As an example, let us consider the Type-II two Higgs doublet model~(2HDM)\cite{Branco:2011iw} 
where  an additional $Z_2$ symmetry is employed
so that one of the doublets~($\phi_2$) couples only to the $T_3=+\frac12$ fermions and the other
doublet~($\phi_1$) couples only to the $T_3=-\frac12$ fermions, thereby ensuring the absence
of any flavor changing neutral current~(FCNC) mediated by neutral scalars. 
We follow the convention of Ref.~\cite{Branco:2011iw} for the parameters
 of the potential as well as the couplings of the scalars
to the fermions and gauge bosons, but extend
this popular framework by adding one complete extra chiral generation of leptons and quarks. 
Note that we have introduced a right-handed neutrino for the fourth generation, and 
made it heavy enough to have no significant impact in collider as well as astrophysical experiments.
 We assume that $(B-L)$ symmetry remains exact in the
Lagrangian and thus consider the neutrinos as Dirac particles. The lightest CP-even
scalar~($h$) in the spectrum of the physical particles is then identified with the
$125$~GeV resonance observed at the LHC. Denoting the ratio of the vacuum expectation
values~(VEVs) of the two scalar doublets by $\tan\beta = v_2/v_1$, we can
write the Higgs coupling modification factors as \cite{Branco:2011iw}
\begin{subequations}
\label{e:ws2}
\begin{eqnarray}
\kappa_V &=& \sin(\beta-\alpha)\,, \quad (V=W,Z) \\
\kappa_u &=& \sin(\beta-\alpha)+\cot\beta\cos(\beta-\alpha)\,, \quad ({\rm for~ up~ type~ quarks})
\label{e:ku} \\
\kappa_d &=& \sin(\beta-\alpha)-\tan\beta\cos(\beta-\alpha)\,, \quad ({\rm for~ down~ type~ quarks~ and ~charged ~leptons})
\label{e:kd}
\end{eqnarray}
\end{subequations}
where $\alpha$ is the mixing angle that takes us from the Lagrangian basis to the
physical basis in the CP-even scalar sector. Evidently, the condition
\begin{eqnarray}
\label{e:ws2HDM}
\cos(\beta-\alpha) = \frac{2}{\tan\beta}\,, \quad {\rm with,} ~\tan\beta \gg 2
\end{eqnarray}
leads to the desired wrong sign limit of \Eqn{e:ws}\cite{Ferreira:2014qda,Fontes:2014tga,
Ferreira:2014dya,Biswas:2015zgk,Ferreira:2014naa}. We also note that by demanding
only $\kappa_u=-\kappa_d = 1$, one obtains from \Eqs{e:ku}{e:kd} the condition,
\begin{eqnarray}
\cos(\beta-\alpha) = \sin2\beta
\end{eqnarray} 
which reduces to \Eqn{e:ws2HDM} in the large $\tan\beta$ limit.

\begin{figure}[htbp!]
\begin{minipage}{0.46\textwidth}
\centerline{\includegraphics[width=8cm,height=6cm]{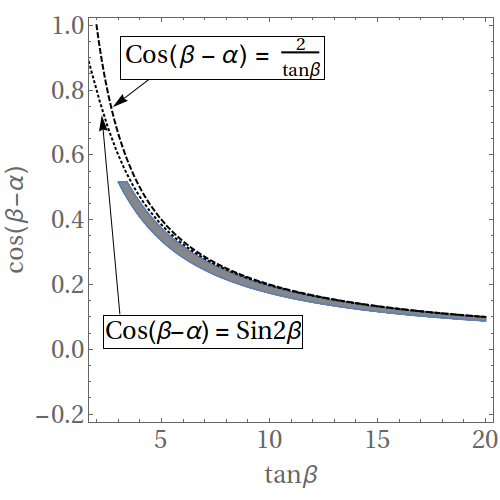}}
\caption{\em The gray shaded region, based on the benchmark given in Eq.\ (\ref{e:bench}), 
 is allowed at 95\% CL from the current data for Higgs signal 
strengths\cite{Khachatryan:2016vau},
  and $\Delta S$, $\Delta T$
for the benchmark of \Eqn{e:bench}. 
The dashed and dotted lines represent the contours for $\cos(\beta-\alpha)=2/\tan\beta$
and $\cos(\beta-\alpha)=\sin2\beta$ respectively.}
  \label{f:ws}
\end{minipage}
\hfill
\begin{minipage}{0.46\textwidth}
\centerline{\includegraphics[width=8cm,height=6cm]{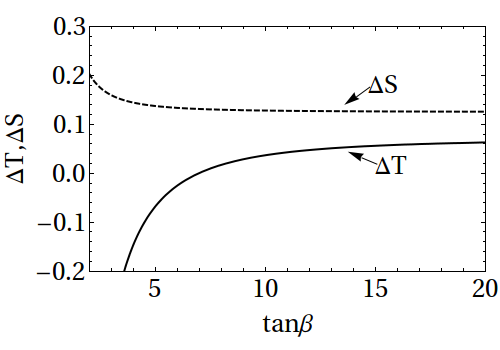}}
\caption{\em Variation of the NP contributions to the $S$ and $T$ parameters with $\tan\beta$ 
assuming the validity of \Eqn{e:ws2}. This plot has been obtained for the benchmark values of \Eqn{e:bench}.}
\label{f:st}
\end{minipage}
\end{figure}

To demonstrate our proposition explicitly, we try to find the allowed region in the
$\tan\beta$~-~$\cos(\beta-\alpha)$ plane so that the Higgs signal strengths into different
production and decay channels remain within the experimental limits\cite{Khachatryan:2016vau}.\footnote{
For $h\to\gamma\gamma$, we have also taken into account the effect of the charged
scalar. However, as
has been argued in Ref.\cite{Bhattacharyya:2014oka}, the charged scalar can be made
to decouple by suitable tuning of the soft breaking mass parameter in scalar potential.}
We display our result in Fig.~\ref{f:ws}. 
This plot shows the allowed parameter space in terms of the two free parameters of the
model, namely, $\cos(\beta-\alpha)$ and $\tan\beta$, for the benchmark point
\begin{eqnarray}
&& m_{t'}= 550~{\rm GeV} \,, \, m_{b'}= 510~{\rm GeV} \,, \, m_{\tau'}= 400~{\rm GeV} \,, \,
m_{\nu'}= 200~{\rm GeV} \,, \nonumber \\
&& m_{H}= 400~{\rm GeV} \,, \, m_{A}= 810~{\rm GeV} \,, \, m_{H+}= 600~{\rm GeV} \,.
\label{e:bench}
\end{eqnarray}
For the parameter space displayed in Fig.~\ref{f:ws}, the BSM effects to the oblique parameters $S$ and $T$,
taking into account both fermionic \cite{Kribs:2007nz,Dighe:2012dz} and scalar \cite{Grimus:2007if,
Grimus:2008nb} contributions, 
are within the experimental limits given by \cite{Olive:2016xmw}\footnote{
If one takes the $1\sigma$ confidence limit on $\Delta S$ and $\Delta T$, the bound on $\tan\beta$
gets modified slightly
along the dotted line in Fig.\ \ref{f:ws} to $\tan\beta \gtrsim 4$.}
\begin{eqnarray}
\Delta S = 0.05\pm 0.10 \,, \qquad \Delta T = 0.08\pm 0.12 \,.
\label{e:stlims}
\end{eqnarray}
In the limit
$\tan\beta \gg 2$ or equivalently $\cos(\beta-\alpha)\to 0$, one gets~(see Fig.~\ref{f:st})
\begin{eqnarray}
\Delta S \approx 0.12 \,, \qquad \Delta T \approx 0.07 \,.
\end{eqnarray}
We have checked that all theoretical constraints on the potential,
{\it e.g.}, perturbative unitarity and stability, are satisfied for such a benchmark point.
One may wonder whether such a benchmark point would not be in conflict with the direct search 
limits from the LHC. We discuss later how such bounds can possibly be evaded. At the same time 
higher mass values for the fourth generation fermions satisfying the $\Delta S$ and 
$\Delta T$ constraints would have been equally good for us.
 
As explained earlier, the mass splitting between the quarks does not affect the cancellation as the 
production amplitude in the ggF channel becomes 
insensitive to the precise mass values at the heavy mass limit; the same applies for the leptons 
while computing the $h\to \gamma \gamma$ amplitude. 

Note that a large splitting between $m_H$ and $m_A$ contributes to the electroweak
$T$-parameter with a negative sign\cite{Branco:2011iw,Bhattacharyya:2013rya,Das:2015qva,Bhattacharyya:2015nca}.
This fact can be used to allow for much larger mass splittings between the components of fourth generation fermion doublets
compared to that in the case of SM4\cite{Kribs:2007nz}.
 For the benchmark of \Eqn{e:bench} we find from Fig.~\ref{f:ws}
that the model automatically  tends to the wrong sign limit, vindicating our assertion.
We have also checked that the nature of the plot does not crucially depend on our choice
of the benchmark as long as all the nonstandard masses are considerably heavier than
$125$~GeV.

%
%

\paragraph{Caveat emptor:}
Lower bounds on the fourth generation quark masses ($m_{q'}\gtrsim 700$~GeV) have been placed
from the direct searches at the LHC\cite{Aad:2015tba,Chatrchyan:2012fp}. These bounds, however,
crucially depend on the assumption that the lighter of the fourth generation quarks~($q'$) decays into a light quark
from the first three generations accompanied by a $W$-boson. But these decay modes can become
subdominant if $V_{i4},V_{4i}\approx 0$, ($i=1,2,3$) in the $4\times4$ CKM matrix. In this case,
$q'$ becomes quasi-stable and, in our toy model in the framework of a Type~II 2HDM, it mainly
decays through the loop-induced neutral currents. In BSM scenarios with tree level FCNC,
where $t'\to th$ is the dominant decay mode for $t'$, the bound on the fourth generation mass
can be relaxed up to $m_{t'}>350$~GeV\cite{Geller:2012wx}. 
The bound of $m_{t'}\sim 850$ GeV for $t'$ decaying through neutral-current channels,
 as recently found by the ATLAS collaboration \cite{ATLAS:2017lvm},
 can be evaded if there are some exotic channels present.
 In addition to the direct searches,
there are also constraints arising from tree-level unitarity of the scattering amplitudes for $2\to 2$
processes of the type $\bar{f}_1f_2 \to \bar{f}_3f_4$~($f_i$ denotes a generic fermion). The
unitarity bound for a fourth generation quark is $m_{q'}\lesssim 550$~GeV\cite{Chanowitz:1978mv,Dighe:2012dz} 
whereas that for
a leptonic fourth generation is $m_{\ell'}\lesssim 1.2$~TeV\cite{Chanowitz:1978mv}. All these considerations together
justifies our choice of the benchmark in \Eqn{e:bench}.

In the moderate $\tan\beta$ region as shown in Fig.\ \ref{f:ws}, one should take into account the bound 
coming from the decay $H\to\tau^+\tau^-$\cite{ATLAS:2017mpg}. For $\tan\beta\sim 10$, 
BR($H\to\tau^+\tau^-)\sim {\cal O}(10^{-2})$ for a type~II 2HDM. In fact, the ATLAS bound, when
translated to the MSSM case, gives $m_{H/A}\gtrsim 250$~GeV for $\tan\beta\sim 10$\cite{ATLAS:2017mpg}. This constraint
can be diluted further if we allow $H$ to decay invisibly into a pair of fourth generation neutrinos.

At this point, let us note that there should be some new dynamics not much above the scale of the new fermion masses. 
Such new dynamics will, in all probability,  alter the bounds arising from
tree-unitarity. The justification for such new dynamics comes from the fact that
the presence of extra heavy
chiral fermions will usually give large negative contributions to the evolution of the scalar
quartic couplings\cite{Kribs:2007nz}. This can potentially render the vacuum unstable very quickly after
the effects of heavy fourth generation set in. This problem can possibly be diluted by
considering a metastable vacuum instead of an absolutely stable one\cite{Kribs:2007nz} and/or adding extra
singlets to the scalar potential, which can give positive contribution to the concerned renormalization group 
equations.

In passing, we also note that,
while such a fourth generation can remain hidden in single production of the Higgs boson, it should 
show up if one considers the double Higgs production, $gg \to hh$, because the new amplitudes will 
then add up instead of cancelling each other. Thus, a significant enhancement of the $gg\to hh$ 
rate may be taken as a possible signature of such an extension of the SM. Non-observation of such an enhancement 
will similarly help to rule the model out.

\paragraph{Summary:}
To summarize, we have presented a general recipe for resurrecting one, or more, sequential
fermion generations in the precision Higgs era ushered by the LHC. Admittedly, we did not provide a complete
model which solves {\em all} the problems faced by a heavy chiral generation.
Nevertheless, we have demonstrated how, in a toy scenario based on a type~II 2HDM,
the sequential fourth generation can overcome its biggest threat, {\it viz}, the consistency
of LHC Higgs data with the SM expectations. Thus our toy model can be taken as a
constituent part of a more elaborate framework which can address all the other issues related
to  such extra chiral generations like the loss of unitarity or the stability of the electroweak
vacuum way above the TeV scale. In anticipation that it will be long before the LHC starts
probing the sign of the bottom quark Yukawa coupling\cite{Modak:2016cdm}, a sequential fourth generation can
remain hidden in the wrong sign limit for many years to come. Hopefully this
unconventional possibility can rekindle the interest in the study of extra fermionic
generations.

\paragraph{Acknowledgements:} AK thanks the Science and Engineering Research Board (SERB), Government of India, 
for an extramural research project. 


\bibliographystyle{JHEP} 
\bibliography{4th_gen.bib}
\end{document}